\documentstyle[aps,prl]{revtex}
\input{epsf}
\begin{document}
\twocolumn[\hsize\textwidth\columnwidth\hsize\csname@twocolumnfalse\endcsname
\bibliographystyle{unsrt}

\draft
\title{Universal Scaling of Wave Propagation Failure in
Arrays of Coupled Nonlinear Cells}
\author{Konstantin Kladko, Igor Mitkov, and A. R. Bishop}
\address{
Theoretical Division and Center for Nonlinear Studies\\
Los Alamos National Laboratory, Los Alamos,
NM 87545, USA 
}
\date{\today}
\maketitle
\begin{abstract}
We study the onset of the
propagation failure
of wave fronts in
systems of coupled cells. We introduce a new method to
analyze the scaling of the critical external field at
which fronts cease to propagate, as a function of intercellular coupling.
We find the universal scaling of the field
throughout the range of couplings,
and show that the field becomes exponentially small for large
couplings. Our method is generic and applicable to a wide
class of cellular dynamics in chemical, biological, and
engineering systems. We confirm our results by direct numerical
simulations.
\end{abstract}


\narrowtext
\vskip1pc]

The impact of discreteness on the propagation of phase fronts in
biophysical, chemical, and engineering systems has been intensively
studied during the last decade. Among the diverse examples are
calcium release waves in living cells~\cite{calcium},
reaction fronts in chains of
coupled chemical reactors~\cite{reactors,reactors1},
arrays of coupled diode resonators~\cite{locher,ditto},
and discontinuous propagation of action potential in cardiac
tissue~\cite{cardiac,munuzuri1,keener}.
All these disparate systems share a common phenomenon of
wave front {\it propagation failure}, independently of
specific details of each system.
Recently this effect has drawn considerable
attention (see, {\it e.g.}, Refs.~\cite{reactors,munuzuri}).
Numerous experimental evidences
show that the propagation failure occurs
at finite values of the coupling strength (a {\it critical
coupling})~\cite{reactors,locher,cardiac}.
This is contrary to continuous systems, where wave fronts propagate
for arbitrary couplings~\cite{ch}.
A challenging problem is to establish the universal properties
of the critical coupling; this is crucial for making predictions
of qualitatively different regimes of system dynamics.

In this Letter we consider the universal behavior of
phase separation fronts in one-dimensional
nonlinear discrete systems in an external field.
We study the propagation failure transition for a class of
simple dynamical models describing experimental observations in
arrays of coupled nonlinear cells, such as
chains of bistable chemical reactors~\cite{reactors,reactors1},
systems of cardiac cells~\cite{keener}, {\it etc}.
A new analytical method is presented to study generic properties
of the critical external field of the transition.
We find, using this method, how the critical field scales
with the intrachain coupling.
This method is applicable for a wide range of the couplings.
We confirm our analytical predictions by direct numerical simulations
of the full system.
Our model in general is given by the following set of
coupled nonlinear equations:

\begin{equation}
\gamma \frac{d u_n}{dt}  \;=\; C (u_{n+1}+u_{n-1}-2 u_n)
- \frac{\partial G(u_n\,,E)}{\partial u_n}\;.
\label{1}
\end{equation}
Here $u_n$ is the order parameter at the $n$-th site, $\gamma$ is the
damping coefficient, $C$ is the coupling
constant, and $G(u_n,E)$ is the onsite potential,
where $E$ is the applied field.
The potential has at least two minima
separated by a barrier $u_B\,$. The external field $E$
is responsible for the energy difference between the minima.
This provides for one globally stable and one metastable minimum,
$u=u_+$ and $u=u_-\,$, respectively. Phase fronts connect the two
minima and tend to propagate, to increase the size of the
energetically more favorable phase, $u_+\,$.

\begin{figure}[h]
\hspace{-1.5cm}
\rightline{ \epsfxsize = 7.0cm \epsffile{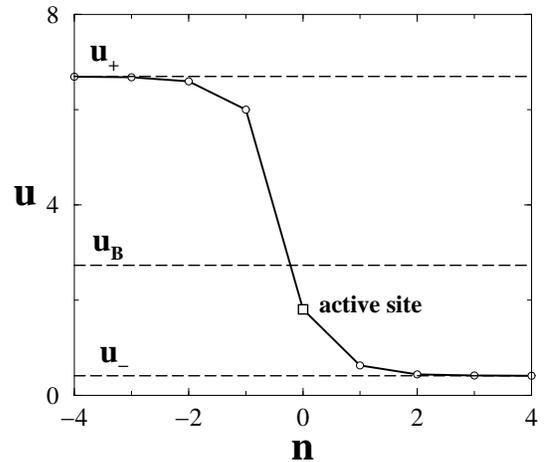}}
\caption{
Stationary front near the propagation failure transition
in Eqs.~(\protect\ref{SG}).
Parameters are $\beta = 0.19\,,\,E=0.4\,$.
Solid line represents the front solution. Dashed lines
correspond to the minima $u_+$ and $u_-$ and the barrier
$u_B\,$.
\label{fig0}}
\end{figure}

The mechanism of front propagation is manifested in the competition
between the system discreteness and the driving field $E\,$.
This competition gives rise to the propagation failure
at the critical field $E_c$, which depends upon the
intercellular coupling~\cite{munuzuri}. For $E>E_c\,$, the front
propagates at the velocity $V$ (see, {\it e.g.}, Ref.~\cite{mitkov})
vanishing at the transition point, $E=E_c\,$.

The results presented in this Letter are generic and apply to
a wide class of systems, independently of the details, as long
as the potential $G(u,E)$ has the bistable structure described
above~\cite{note}.
We start our analysis with the sine-Gordon
potential, $G = K(1-\cos u-E u)\,$, with
the factor $K$ being the potential amplitude.
This potential was chosen in order to address the systems with
phase dynamics, where the order parameter possesses a natural
periodicity, such as arrays of Josephson junctions~\cite{flach}.
The dimensionless dynamics in this case are given by
\begin{equation}
\frac{d u_n}{dt}  \;=\; \beta (u_{n+1}+u_{n-1}-2 u_n)
+ E - \sin u_n\;,
\label{SG}
\end{equation}
where time is rescaled as $t \rightarrow tK/\gamma$ and
the dimensionless coupling is $\beta=C/K\,$.
Our consideration is focused on the elementary
fronts connecting two nearest minima out of the infinite set of
the potential minima. The system is invariant with respect
to the shift $u \rightarrow u + 2\pi\,$, so we choose
without loss of generality
$u_- = \arcsin E\,,\, u_+ = u_- + 2\pi\,,\,u_B=\pi-\arcsin E\,$.
This makes the dynamics effectively bistable.

We introduce a novel analysis to find the universal
dependence of the critical field $E_c$ on the coupling $\beta\,$.
For not too large $\beta\,$, the results are in a good quantitative
agreement with the corresponding asymptotic description.
Finally we address a bistable fourth degree polynomial potential
with the corresponding force in Eq.~(\ref{1}),
$\,-\partial G/\partial u = -u(u-1)(u-1/2+E)\,$.
Such a potential is applied, {\it e.g.}, to describe the propagation
failure in arrays of chemical reactors~\cite{reactors,reactors1}
and coupled diode resonators~\cite{locher,ditto}.
This illustrates the applicability of our approach to generic
potentials. The obtained analytical results for both potentials
are justified by the numerical simulations of the full systems.

For large dimensionless coupling, $\beta \gg 1\,$, Eqs.~(\ref{SG})
approach the continuous regime described by the
overdamped sine-Gordon equation, $u_t=\beta u_{xx} - \sin u + E\,$.
Here $x$ is a spatial coordinate standing for the continuous site number.
This equation possesses a front solution, which propagates at nonzero
velocity, for any driving field $E\,$. For small field, the front
has the form $u \approx 4 \arctan \left[ \exp(z/\sqrt{\beta}) \right]\,$,
with the traveling wave coordinate $z=x-Vt\,$ and velocity
$V \sim E \sqrt{\beta}$~\cite{mihailov}. Therefore,
the critical field $E_c$ vanishes in the continuous regime.
For finite $\beta\,$, however, the external driving can be balanced
by the effect of discreteness, thus leading to a propagation failure
at a finite value of $E=E_c(\beta)\,$.
Note that for a wide class of bistable
systems, there is a global instability at $E = E_{gl}\,$, above which
the potential minima $u_{\pm}$ (or one of them) cease to exist.
Typically the propagation failure occurs for $E_c < E_{gl}\,$.
In particular, for Eqs.~(\ref{SG}) one has $E_{gl} = 1\,$.

In order to study the onset of the propagation failure, one has to
consider the stationary case of Eqs.~(\ref{SG}), {\it i.e.},
$\partial u_n/\partial t = 0\,$ (Fig~\ref{fig0}).
We denote the site closest
to the barrier separating two potential minima, as the
``front site'' and assign to it number $n=0\,$.

Before developing the main approach of the present study,
we briefly present our results obtained by means of
{\it single-active-site theory}, valid for the
discrete regime of not too large coupling $\beta\,$.
This theory is based on the fact that, for such $\beta\,$,
only the front site, $n=0\,$, experiences
nonlinearity, while
all the other sites are close enough to either $u_+$ or $u_-\,$
to be in a linear regime, see Fig~\ref{fig0}.
Solving the linearized version of
Eqs.~(\ref{SG}) in the stationary case, one obtains
$u_n=  \Delta  + A \exp({\lambda n})$ and
$u_n= 2 \pi + \Delta  +B \exp({-\lambda n})\,$, for
$n\;\le\;0$ and $n\ge\;0\,$, respectively.
Here $\Delta  =\arcsin{E}\,$,
$\lambda = \mbox{arccosh}[\,1/(2\beta)+\cos\Delta\,]\,$.
Matching these at $n=0$ leads to
$A=2 \pi + B\,$, which after substitution in the full nonlinear
equation for $u_0\,$, yields
\begin{equation}
- 2\beta\,\left(\pi+B)(e^{-\lambda}-1\right)
+ \sin{(B+ \Delta )}=E \;.
\label{bif}
\end{equation}
For $E<E_c\,$, Eq.~(\ref{bif}) possesses two solutions,
corresponding to stable and unstable fronts in the
original problem (\ref{SG}). At the bifurcation point
$E=E_c$ these solutions merge and disappear,
so that for $E>E_c$ no stationary solution to Eqs.~(\ref{SG})
exists, and the front propagates. This implies that
at the bifurcation point the derivative of the left hand
side of Eq.~(\ref{bif}) with respect to $B$ must equal zero.
Then we obtain, after some calculations, an approximate expression
for $E_c\,$:
\begin{eqnarray}
E_c \approx &&\frac{\sqrt{2\sqrt{1+4\beta}-4\beta -1}
-f(\beta)\arccos{[f(\beta)]}} {1+f(\beta)}\;,
\label{asympt}\\
&&\mbox{where}\;\;\;\;
f(\beta)=2\left[1-\frac{\beta+1}{1+2\beta+\sqrt{1+4\beta}}\right]\;.
\nonumber
\end{eqnarray}

\begin{figure}[h]
\hspace{-0.7cm}
\rightline{ \epsfxsize = 8.0cm \epsffile{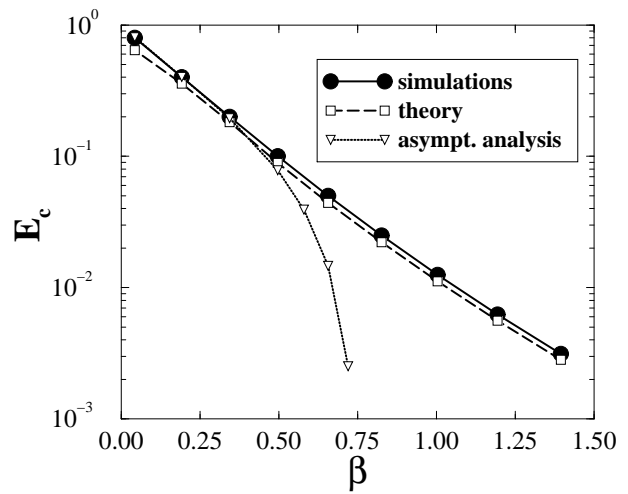}}
\caption{
Critical curve of the propagation failure transition
for the sine-Gordon potential.
Predictions of our general method and single-active-site
theory are compared with numerical simulations
of system~(\protect\ref{SG}).
The system length is $L = 100$, and time step of the simulations
$dt = 0.0001$.
\label{fig1}}
\end{figure}

The graph of $E_c(\beta)$ given by (\ref{asympt}) is
shown in Fig.~\ref{fig1} (dotted line). We see from the figure that
it agrees well with the results of numerical simulations
of the full system~(\ref{SG}), for small to moderate
values of $\beta$ ($\beta \lesssim 0.5$). This makes the
single-active-site theory substantially more useful than a regular
small $\beta$ perturbation theory, which works only for much
smaller $\beta$ ($\lesssim 0.1$).

Although the single-active-site theory gives reasonable predictions
for the propagation failure transition for moderate values
of the intrachain coupling $\beta\,$, it does not provide a universal
scaling of the critical field $E_c$ with $\beta\,$ (see Fig.~\ref{fig1}).
This motivates developing a
{\it general theory} to describe the phenomenon
of the propagation failure
throughout the range of
$\beta$. Such a theory is presented below.

A stationary front in Eq.~(\ref{SG}) can be obtained as an extremum
of the free energy
\begin{equation}
{\mathcal E} = \sum_n \,[\,\frac{\beta}{2}(u_{n+1}-u_n)^2
+ 1-\cos u_n-Eu_n\,]\;.
\label{energy}
\end{equation}
For the case of zero field $E$, the front is always stationary, and
no transition exists. In this case we have for the front,
taking into account the effect of discreteness
(see also \cite{flachkladko})
\begin{eqnarray}
&&u_n = w(n) - \frac{\sin w(n)}{12\beta}\;,
\label{4}\\
&&w(x) = - 2\,{\rm arctan}\left[ b\,
\sinh^{-1} \left( \frac{b\,x}{\sqrt{\beta}} \right)
 \right] \;, \;\;\;\;\;\; x\leq 0 \;,
\nonumber\\
&&w(x) = 2\pi - w(-x)\;,
\;\;\;\;\;\;\;\;\;\;\;\;\;\;\;\;\;
\;\;\;\;\;\;\;\;\;\;\;\;\;\;\;\,
x\geq 0\;,
\nonumber
\end{eqnarray}
with the coefficient $b = [\,1-1/(12\beta)\,]^{1/2}$.

At zero field $E\,$, the free energy ${\mathcal E}$ (\ref{energy})
possesses an infinite set of minima of equal depths,
separated by barriers. Each of these minima corresponds
to a stable stationary front and differs from the other
fronts by a shift on an integer number of sites.
Each barrier corresponds to an unstable front.
When the external field $E$ is applied, the set of
minima is tilted, so that the difference between their
depths is determined by the field value. As the field
increases, each of the minima approaches its
adjacent barrier (down the energy landscape), until they
finally merge, at $E=E_c\,$. For $E>E_c\,$, the energy
has no extrema left, and therefore the fronts propagate
without limit.

Our method is based on choosing profiles $\{u_n\}$ that
can be parameterized
by an effective coordinate $\alpha\,$, and replacing the argument $n$ of
function $w(n)$ in (\ref{4}) with $n+\alpha\,$.
We show below that this choice allows us to find the critical
field $E_c\,$. It can be demonstrated that, for $E=0\,$, the points
$\alpha=1/2+m$ ($m$ is an integer)
correspond to stable fronts, and $\alpha=m\,$, to unstable fronts.
The former and the latter are energy minima and maxima,
respectively, satisfying $d{\mathcal E}/d\alpha = 0\,$.
Note that the chosen parameterization by $\alpha$ may be interpreted
as a continuous shift of a profile $\{u_n\}\,$.

Our goal now is to evaluate the free energy landscape
${\mathcal E}(\alpha)$ given by (\ref{energy}),
for nonzero field $E\,$. When $E=E_c$ the minima of ${\mathcal E}(\alpha)$
merge with their adjacent maxima at the inflection points
$d^2{\mathcal E}/d\alpha^2=0\,$. We have substituted the fronts (\ref{4})
with $w(n+\alpha)$ into energy (\ref{energy})
and found $E_c$ for various $\beta$'s,
using {\it Mathematica} software.
The result is shown in Fig.~\ref{fig1} (dashed line).

In order to find the analytical form of ${\mathcal E(\alpha)}$, we use
the Poisson summation formula for infinite series:
\begin{equation}
\sum_{-\infty}^{\infty} F(n) =
\int_{ -\infty}^{\infty} F(x) \left[\,1 +
2\sum_{k=1}^{\infty} \cos(2\pi kx)\,\right] dx\;,
\label{6}
\end{equation}
where $F(n)$ is the $n$th element of the series in (\ref{energy}).
One finds that the integral $\int F(x)dx \approx E\alpha\,$, up to
a term independent of $\alpha$. Then, after closing the integration contour
in the complex plane, we see that
$k>1$ terms in (\ref{6}) are exponentially small, compared to the $k=1$ term.
This dominant term can be evaluated approximately to give the
following expression for ${\mathcal E}(\alpha)$
\begin{equation}
{\mathcal E}(\alpha) = 2\pi E \alpha +
\Omega\beta\cos(2\pi\alpha)
\exp\left[-\frac{\pi^2\beta}{\sqrt{\beta-1/12}}\right]\;.
\label{analytic}
\end{equation}
The factor $\Omega$ is a constant and can be found by comparison of
(\ref{analytic}) with the results obtained by {\it Mathematica}
(see Fig.~\ref{fig1}), which yields $\Omega\approx 340\,$.
Then, using the conditions
$d{\mathcal E}/d\alpha = d^2{\mathcal E}/d\alpha^2 = 0\,$,
we find 
\begin{equation}
\alpha = \frac{1}{4}\;,\;\;\;\;\;\;\;\;\;
E_c = \Omega\beta\,\exp\left[-\frac{\pi^2\beta}{\sqrt{\beta-1/12}}\right]\;.
\label{EC}
\end{equation}
The general scaling $E_c(\beta)$ of the propagation failure
transition given by (\ref{EC}), virtually coincides with
the one obtained by {\it Mathematica}, shown in
Fig.~\ref{fig1}~\cite{note1}. In the continuous regime of
large $\beta$ the result~(\ref{EC}) has the
following simple asymptotic form
$E_c=\Omega\beta\exp[-\pi^2\sqrt{\beta}]\,$.
This implies that the critical field $E_c$ decays
exponentially for large couplings $\beta\,$.

To confirm our general theory,
we have performed numerical simulations of the full system~(\ref{SG}).
We have used the implicit second order integration method,
in order to obtain the transition line $E_c(\beta)\,$.
The results are presented in Fig.~\ref{fig1} (solid line).
We see in the figure a good quantitative agreement of
our analytical predictions with the results of the
simulations.

\begin{figure}[h]
\hspace{-0.7cm}
\rightline{ \epsfxsize = 8.0cm \epsffile{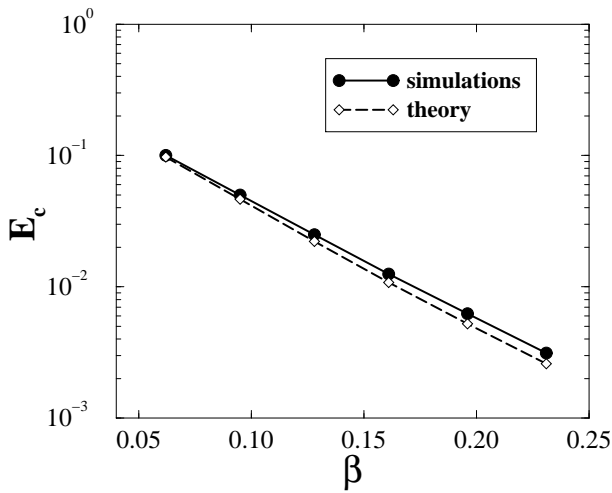}}
\caption{
Critical curve of the propagation failure transition
for the fourth degree polynomial potential.
The system length and and time step of the simulations
are the same as in Fig.~\protect\ref{fig1}.
\label{fig2}}
\end{figure}

To address the propagation failure in arrays of chemical
reactors~\cite{reactors,reactors1} and diode
resonators~\cite{locher,ditto}, we turn to a fourth degree polynomial
potential leading to a cubic force in Eqs.~(\ref{1}):
$\,-\partial G/\partial u = -u(u-1)(u-1/2+E)\,$.
The globally stable and metastable minima for this potential
are $u_+=1$ and $u_-=0\,$, respectively; the barrier $u_B=1/2-E\,$.
Applying our general theory
developed above, we find the propagation failure transition
line $E_c(\beta)\,$. In Fig.~\ref{fig2} we show the results
obtained by {\it Mathematica} (dashed line).
We see again that the theory is confirmed by numerical simulations
of the full system (solid line). The analytic expression of
$E_c(\beta)$ analogous to~(\ref{EC}) is cumbersome~\cite{kladko},
so in this Letter we only give its asymptotic form for large
$\beta\,$: $E_c = \Omega\beta\exp(-\eta\sqrt{\beta})\,$.
This form has the same structure as the asymptotics of critical
curve~(\ref{EC}) for sine-Gordon potential,
which provides a persuasive argument that the proposed
approach is generic, independent of the specific potential. The values
of $\Omega$ and $\eta$ are found to be
$\Omega \approx 429\,,\,\eta\approx 21.9\,$.

To conclude, we have introduced a general method to analyze
the transition of front propagation failure in arrays of coupled
nonlinear cells. Using this method we have determined,
for bistable dynamics of the cells,
how the critical external field of the transition scales
with intrachain coupling. To demonstrate the generality
of the new method, we have carried out our analysis
for two different effectively bistable potentials:
({\it i}) the sine-Gordon
potential, useful {\it e.g.}, for describing arrays of
Josephson junctions~\cite{flach}; and ({\it ii}) a fourth degree
polynomial potential applicable to
chains of chemical reactors~\cite{reactors,reactors1}
and coupled diode resonators~\cite{locher,ditto}.
Our theoretical predictions have been confirmed
by numerical simulations of the full systems.
After this study of the propagation failure problem
in the one-component systems~(\ref{1}), the following question
arises: what is the mechanism of this phenomenon in systems
described by more complicated dynamics? For example,
the ``single-pool-model'' for intracellular calcium
waves~\cite{atri} is represented by two dynamic components:
the calcium concentration and the fraction of vacant ionic
channels. An important and virtually unexplored issue
is how the effect of discreteness of ionic channels
determines the onset of the propagation failure of calcium waves.
We expect that our techniques can provide
new insights into this and other types of 
more complicated chains of coupled nonlinear elements
in biophysics, chemistry, and engineering.
A further challenging direction of study here is to extend
our analysis of the propagation failure to two- and
three-dimensional systems. This would allow one to understand
the effect of discreteness on the dynamics and stability
of such structures as spiral and scroll waves in cardiac
tissue~\cite{munuzuri,osipov,mitkov1}.

We thank John Pearson for fruitful discussions and
valuable advice. This work was supported by the Department
of Energy under contract W-7405-ENG-36.


\begin{references}
\vspace{-1.2cm}

\bibitem{calcium}
A. E. Bugrim, A. M. Zhabotinsky, and I. R. Epstein,
Biophys.\ J. {\bf 73}, 2897 (1997);
J.\ Keizer, G. D.\ Smith, S.\ Ponce Dawson,
and J. E. Pearson, Biophys.\ J. {\bf 75}, 595 (1998);
S. P. Dawson, J. Keizer, and J. E. Pearson, Proc. Natl. Ac. Sci. U.S.A.,
{\bf 96}, 6060 (1999).

\bibitem{reactors} J. P. Laplante and T. Erneux,
J. Phys. Chem. {\bf 96}, 4931 (1992); Physica (Amsterdam)
{\bf 188A}, 89 (1992).

\bibitem{reactors1} V. Booth, T. Erneux, and J. P. Laplante,
J. Phys. Chem., {\bf 98}, 6537 (1994).

\bibitem{locher}
M. Locher, G. A. Johnson, and E. R. Hunt, \prl {\bf 77}, 4698 (1996);
M. Locher, D. Cigna, and E. R. Hunt,
\prl {\bf 80}, 5212 (1998).

\bibitem{ditto} 
J. F. Lindner, S. Chandramouli, A. R. Bulsara,
M. Locher, and W. L. Ditto, \prl {\bf 81}, 5048 (1998).

\bibitem{cardiac} W. C. Cole, J. B. Picone, and N. Sperelakis,
Biophys. J. {\bf 53}, 809 (1988).

\bibitem{munuzuri1} M. deCastro, E. Hofer, A. P. Munuzuri,
M. Gomez-Gesteira, G. Plank, I. Schafferhofer, V. Perez-Munuzuri,
and V. Perez-Villar,
\pre, {\bf 59}, 5962 (1999).

\bibitem{keener} J. P. Keener, J. Theor. Biol.
{\bf 148}, 49 (1991).

\bibitem{munuzuri} A. P. Munuzuri, V. Perez-Munuzuri, M. Gomez-Gesteira,
L. O. Chua, and V. Perez-Villar,
Int. J. Bifurcation Chaos {\bf 5}, 17 (1995).

\bibitem{ch} M. Cross and P. C. Hohenberg, \rmp {\bf 65} (1993).

\bibitem{mitkov} I. Mitkov, K. Kladko, and J. E. Pearson,
\prl {\bf 81}, 5453 (1998).

\bibitem{note}
Although (\protect\ref{1}) is an overdamped equation, our
method also holds in the underdamped case, with $d^2 u_n/dt^2$
present, since it is based on the stability properties of
stationary solutions.

\bibitem{flach} S. Flach and M. Spicci, J. Phys. C {\bf 11},
321 (1999).

\bibitem{mihailov} A. S. Mikhailov, {\it Foundations of Synergetics I:
Distributed Active Systems} (Springer-Verlag, Berlin, 1990).

\bibitem{flachkladko} S. Flach and K. Kladko,
\pre {\bf 54}, 2912 (1996).

\bibitem{note1} The theory given by (\protect\ref{EC}) is applicable
for $\beta\gtrsim 1/12\,$, because of the singularity in the front
(\protect\ref{4}). However, for $\beta\lesssim 1/12\,$, the propagation
failure transition is described well by the
single-active-site theory (\protect\ref{asympt}) (see
dotted line in Fig.~\protect\ref{fig1}).

\bibitem{kladko} K. Kladko, I. Mitkov, and A. R. Bishop,
to be published elsewhere.

\bibitem{atri} A. Atri, J. Amundson, D. Clapham, and J. Sneyd,
Biophys. J. {\bf 65}, 1727 (1993).

\bibitem{osipov}
G. V. Osipov, B. V. Shulgin, and J. J. Collins,
\pre {\bf 58}, 6955 (1988).

\bibitem{mitkov1} I. Aranson  and  I. Mitkov,
\pre {\bf 58}, 4556 (1998).


\end{references}
\end{document}